\documentstyle[epsfig]{mn2e}

\title[GW background from black hole formation]
{Gravitational wave background from Population III black hole
formation}
      { }
\author[J.C.N. de Araujo, O.D. Miranda and O.D. Aguiar]
           {Jos\'e C. N. de Araujo\thanks{jcarlos@das.inpe.br},
            Oswaldo D. Miranda\thanks{Present address: Department
            of Physics, Washington University, Campus Box 1105,
            One Brookings Drive, St. Louis - MO~63130-4899 - USA;
            oswaldo@hbar.wustl.edu}
           and Odylio D. Aguiar\thanks{odylio@das.inpe.br} \\
Instituto Nacional de Pesquisas Espaciais - Divis\~ao de Astrof\'\i sica \\
Av. dos Astronautas 1758, S\~ao Jos\'e dos Campos, 12227-010 SP, Brazil}

\date{Accepted 2001 October 26}

\pagerange{\pageref{firstpage}--\pageref{lastpage}} \pubyear{2001}

\begin{document}

\maketitle

\label{firstpage}

\begin{abstract}
We study the generation of a stochastic gravitational wave (GW)
background produced from a population of core-collapse supernovae,
which form black holes in scenarios of structure formation. We
obtain, for example, that the formation of a population
(Population III) of black holes, in cold dark matter scenarios,
could generate a stochastic GW background with a maximum amplitude
of $h_{\rm BG} \simeq 10^{-24}$ and corresponding closure energy
density of $\Omega_{\rm{GW}}\sim 10^{-7}$, in the frequency band
$\nu_{\rm{obs}} \simeq 30-470\, {\rm Hz}$ (assuming a maximum
efficiency of generation of GWs, namely,  $\varepsilon_{\rm
GW_{\rm max}} = 7\times 10^{-4}$) for stars forming at redshifts
$z\simeq 30-10.$ We show that it will be possible in the future to
detect this isotropic GW background by correlating signals of a
pair of `advanced' LIGO observatories (LIGO III) at a
signal-to-noise ratio of $\simeq 40$. We discuss what
astrophysical information could be obtained from a positive (or
even a negative) detection of such a GW background generated in
scenarios such as those studied here. One of them is the
possibility of obtaining the initial and final redshifts of the
emission period from the observed spectrum of GWs.

\end{abstract}


\section{Introduction}

The detection of gravitational waves (GWs) will open up a new era
in the history of astronomy and transform research in general
relativity into an observational/theoretical study (Schutz 1999).

The detection of GWs will directly verify the predictions of
general relativity theory concerning the existence or not of such
waves, as well as other theories of gravity (Thorne 1987). The
information provided by such waves is completely different when
compared to that provided by electromagnetic waves. GWs carry
detailed information on the coherent bulk motions of matter, such
as in collapsing stellar cores or coherent vibrations of
space-time curvature as produced, for example, by black holes. On
the other hand, electromagnetic waves are usually an incoherent
superposition of emissions from individual atoms, molecules and
charged particles.

There is a host of possible astrophysical sources of GWs: namely,
supernovae, the collapse of a star or star cluster to form a black
hole, inspiral and coalescence of compact binaries, the fall of
stars and black holes into supermassive black holes, rotating
neutron stars, ordinary binary stars, relics of the big bang,
vibrating or colliding of monopoles, cosmic strings and cosmic
bubbles, among others (see, e.g., Thorne 1987, 1996, 1997; Schutz
1996, 1999).

From the theoretical point of view there has been a great effort
to study which are the most promising sources of GWs to be
detected. In particular, the waveforms, the characteristic
frequencies and the number of sources per year that one expects to
observe are questions that have been addressed (see Thorne 1997;
Schultz 1999; Grishchuk et al. 2000 for a review). In a few years,
starting from the observations (waveforms, amplitudes,
polarizations, etc.), it will be possible really to understand how
GW emission is generated by the astrophysical sources.

Because of the fact that GWs are produced by a large variety of
astrophysical sources and cosmological phenomena, it is quite
probable that the Universe is pervaded by a background of such
waves. A variety of binary stars (ordinary, compact or
combinations of them), Population III stars, phase transitions in
the early Universe and cosmic strings are examples of sources that
could produce such putative GW background (Thorne 1987).

As GWs possess a very weak interaction with matter, passing
through it without being disturbed, once detected they can provide
information on the physical conditions from the era in which they
were produced. In principle, it will be possible to get
information from the epoch when the galaxies and stars started to
form and evolve.

Concerning the production of GW backgrounds, it is worth
mentioning that recently Blair \& Ju (1996) and Ferrari, Matarrese
\& Schneider (1999a,b) studied the cosmological GW background
produced by supernovae explosions that took place in the redshift
range $0< z <5$.

On the other hand, from considerations based on the Gunn-Peterson
effect (Gunn \& Peterson 1965), it is widely accepted that the
Universe underwent a reheating (or reionization) phase between the
standard recombination epoch (at $z\sim 1000$) and $z > 5$ (see
Haiman \& Loeb 1997; Loeb \& Barkana 2001 for a review). However,
at what redshift the reionization occurred is still an open
question (Loeb \& Barkana 2001), although recent studies conclude
that it occurred at redshifts in the range $6 < z < 30$
(Venkatesan 2000; Tegmark \& Zaldarriaga 2000; Schmalzing et al.
2000; Loeb \& Barkana 2001). It is worth noting that present and
future cosmic background radiation (CBR) studies can impose some
constraints on the reionization phase of the Universe (see, e.g.,
Tegmark \& Zaldarriaga 2000;  Loeb \& Barkana 2001).

Although different models could possibly explain the reionization
of the Universe, it is widely accepted that most of the
contribution to the reionization is related to the formation and
evolution of pre-galactic objects (sometimes called Population III
objects) at high redshift ($z>10$), such as subgalactic objects
($M<10^{9} {\rm M}_\odot$) and stars formed from them (Loeb \&
Barkana 2001). This putative epoch where the formation of the
Population III objects took place, and where the consequent
reionization and reheating of the Universe occurred, marked the
end of an epoch named `dark age' (see, e.g., Rees 1998). Also,
note that the metalicity of $\sim 10^{-2} Z_{\odot}$ found in
${\rm high}-z$ Ly$\alpha$ forest clouds (Songaila \& Cowie 1996;
Ellison et al. 2000) is consistent with a stellar population
formed at $z
> 5$ (Venkatesan 2000).

The history of the Universe during the formation of the Population
III objects could be investigated in the future with the {\it New
Generation of Space Telescope} ({\it NGST}; Rees 1998) and also,
in principle, with GW observatories. Besides the reheating and
reionization phases, putative Population III stars could produce
GWs, particularly, from the formation of neutron stars and black
holes. Also, after the dark age, the epoch of the first light
could be studied with large radio telescopes such as the Giant
Meter Wavelength Telescope (GMRT) or the Square Kilometer Array
Radio Telescope (SQA; Meiksin 1999).

When the high-mass stars died as supernovae, they left stellar
black holes as remnants. The formation of these stellar Population
III black holes can, in principle, produce a GW background
detectable by GW observatories. It is worth mentioning that a
significant amount of GWs can also be produced during the
formation of neutron stars. However, because this depends on the
equation of state for the neutron star, which is not well defined,
we consider here only the contribution of the black holes. Another
possibility would be the generation of GWs through the so-called
r-mode instability (Anderson 1998), which should be important for
young, hot and rapidly rotating neutron stars, but we leave this
issue for another study, to appear elsewhere.

In the present study we have adopted a stellar generation with a
Salpeter initial mass function (IMF) as well as different stellar
formation epochs. We then discuss what conclusions would be drawn
whether (or not) the stochastic background studied here is
detected by the forthcoming GW observatories such as LIGO and
VIRGO.

The paper is organized as follows. In Section 2 we present the
basic ideas on the collapse of the first clouds and on the
resulting stellar formation; in Section 3 we describe how to
calculate the GW background produced during the formation of black
holes in this scenario; in Section 4 we present and discuss the
numerical results; in Section 5 we consider the detectability of
the putative GW background produced by the Population III black
holes; and finally in Section 6 we present the conclusions.

\section{Population III objects and the first stars}

The current theory of structure formation, based on cold dark
matter (CDM) models, predicts that the first objects to collapse,
the so-called Population III objects or mini-haloes, had a total
mass of $\sim 10^6 M_\odot$ and a formation epoch $z \sim 10 - 50$
(see Tegmark et al. 1997, and references therein). The first
stars, the Population III stars, started forming in these
Population III objects of $\sim 10^6 M_\odot$ and subsequently in
more massive mini-haloes (see Haiman \& Loeb 1997; Venkatesan
2000; Loeb \& Barkana 2001).

As a result of the formation of the first stars, a population of
stellar black holes is formed after the supernova explosions
associated with the high-mass stars. Then, knowing the law of
distribution of stellar masses, it is possible to obtain the
number of stars that explode as supernovae, and so it is possible
to determine the number (or event rate) of stellar black holes
left as remnants.

Thus, to proceed, the distribution function of stellar masses, the
stellar IMF, for the first stars is required. Here the Salpeter
IMF is adopted, namely

\begin{equation}
\phi(m) = A m^{-(1+x)},
\end{equation}

\noindent where $A$ is the normalization constant and $x=1.35$
(our fiducial value). The normalization of the IMF is obtained
through the relation

\begin{equation}
\int_{m_{\rm l}}^{m_{\rm u}} m\phi(m)dm = 1,
\end{equation}

\noindent where we consider $m_{\rm l} = 0.1\, {\rm M}_{\odot}$
and $m_{\rm u}=125\, {\rm M}_{\odot}$. It is worth noting that
some authors argue (see, e.g., Gilmore 2001) that there is
evidence supporting the universality of the IMF, even for the
first stars; on the other hand, other authors (see, e.g., Scalo
1998, among others) argue that the IMF may not be universal. In
particular, the universality of the Salpeter exponent ($x=1.35$)
has been studied by recent evolutionary models for the Magellanic
Clouds (de Freitas-Pacheco 1998). Some models, particularly those
of the Large Magellanic Cloud, take into account constraints on
the star formation history imposed by recent data on
color-magnitude diagrams of field star clouds, showing that a
steeper exponent, $x=2.0$ is necessary to resolve the excessive
production of iron obtained if one takes into account the Salpeter
law ($x=1.35$). Furthermore, concerning the star formation at high
redshift, the IMF could be biased toward high-mass stars, when
compared to the solar neighborhood IMF, as a result of the absence
of metals (Bromm, Coppi \& Larson 1999, 2001).

Then, for the standard IMF, the mass fraction of black holes
produced as remnants of the stellar evolution is

\begin{equation}
f_{\rm BH} = \int_{\rm m_{min}}^{m_{\rm u}}M_{\rm r}\phi(m)dm,
\end{equation}

\noindent where $m_{\rm min}$ is the minimum stellar mass capable
of producing a black hole at the end of its life, and $M_{\rm r}$
is the mass of the remnant black hole. Timmes, Woosley \& Wheaver
(1995) (see also Woosley \& Timmes 1996) obtain, from stellar
evolution calculations, that the minimal progenitor mass to form
black holes is $18\leq m_{\rm min} /{\rm M}_{\odot}\leq 30$
depending on the stellar iron core mass. Thus, we assume that the
minimum mass capable of forming a remnant black hole is $m_{\rm
min}=25\,{\rm M}_{\odot}$. For the remnant, $M_{\rm r}$, we take
$M_{\rm r}=\alpha\, m$, where $m$ is the mass of the progenitor
star and $\alpha = 0.1$ (see, e.g., Ferrari et al. 1999a, b). With
these considerations at hand, the mass fraction of black holes
reads $f_{\rm BH}=6.8\times 10^{-2}\times \alpha \simeq\,
6.8\times 10^{-3}$ for $x=1.35$.

To assess the role of possible IMF variations in our results,
other values of $x$ have also been considered. Besides the
standard IMF, two others have been studied, namely, with $x=0.3$
and $x=1.85$, which yield ten times and one-tenth of the mass
fraction of black holes of the standard IMF, respectively.

It is worth mentioning that stars formed with masses greater than
$8\, {\rm M}_{\odot}$ to $\sim\, 25\,{\rm M}_{\odot}$ also finish
their lives as supernovae. Numerical studies have shown that these
stars leave neutron star remnants, after forming iron cores with
masses near the Chandrasekhar limit (Woosley \& Timmes 1996).
These stars are important for injecting energy into the ambient
medium and regulating the feedback of stellar formation. In the
present paper only the generation of GWs that come from black
holes formation have been studied, and so the progenitors of
interest are stars with masses in the interval $25\leq m/{\rm
M}_{\odot}\leq 125$.

\section{Gravitational wave production}

The GWs can be characterized by their dimensionless amplitude,
$h$, and frequency, $\nu$. The spectral energy density, the flux
of GWs, received on Earth, $F_\nu$, in ${\rm erg}\; {\rm
cm}^{-2}{\rm s}^{-1}{\rm Hz}^{-1}$, is (see, e.g., Douglass \&
Braginsky 1979; Hils, Bender \& Webbink 1990)

\begin{equation}
F_{\nu} = {c^{3}s_{\rm h}\omega_{\rm obs}^{2}\over 16{\rm \pi} G},
\label{fluxa}
\end{equation}

\noindent where $\omega_{\rm obs}=2{\rm \pi} \nu_{\rm{obs}}$, with
$\nu_{\rm{obs}}$ the GW frequency (Hz) observed on Earth, $c$ is
the velocity of light, $G$ is the gravitational constant and
$\sqrt{s_{\rm h}}$ is the strain amplitude of the GW ($\rm
Hz^{-1/2}$).

The stochastic GW background produced by gravitational collapses
that lead to black holes would have a spectral density of the flux
of GWs and strain amplitude also related to the above equation
(\ref{fluxa}). Therefore, in the above equation the strain
amplitude takes into account the star formation history occurring
at the `first light', just after the `dark age' epoch. The strain
amplitude at a given frequency, at the present time, is a
contribution of black holes with different masses at different
redshifts. Thus, the ensemble of black holes formed produces a
background whose characteristic strain amplitude at the present
time is $\sqrt s_{\rm h}$.

On the other hand, the spectral density of the flux can be written
as (Ferrari et al. 1999a,b)

\begin{equation}
F_{\nu}=\int_{z_{\rm {cf}}}^{z_{\rm {ci}}} \int_{m_{\rm
{min}}}^{m_{\rm {u}}} f_{\nu}(\nu_{\rm{obs}}) dR_{\rm BH}(m,z),
\end{equation}

\noindent where $f_{\nu}(\nu_{\rm{obs}})$ is the energy flux per
unit of frequency (in ${\rm erg}\;{\rm cm}^{-2}{\rm Hz}^{-1}$)
produced by the formation of a unique black hole and $dR_{\rm BH}$
is the differential rate of black holes formation.

The above equation takes into account the contribution of
different masses that collapse to form black holes occurring
between redshifts $z_{\rm ci}$ and $z_{\rm cf}$ (beginning and end
of the star formation phase, respectively) that produce a signal
at the same frequency $\nu_{\rm{obs}}$. On the other hand, we can
write $f_{\nu}(\nu_{\rm{obs}})$ (Carr 1980) as

\begin{equation}
f_{\nu}(\nu_{\rm{obs}}) = {{\rm \pi} c^{3}\over 2G}h_{\rm BH}^{2},
\end{equation}

\noindent where $h_{\rm BH}$ is the dimensionless amplitude
produced by the collapse to a black hole of a given star with mass
$m$ that generates at the present time a signal with frequency
$\nu_{\rm{obs}}$. Then, the resulting equation for the spectral
density of the flux is

\begin{equation}
F_{\nu} = {\pi c^{3}\over 2G} \int h_{\rm  BH}^{2}dR.
\end{equation}

\par\noindent From the above equations we obtain for the strain
amplitude

\begin{equation}
s_{\rm h} = {1 \over \nu_{\rm obs}^{2}}\int h_{\rm BH}^{2} dR.
\end{equation}

\par\noindent Thus, the dimensionless amplitude reads

\begin{equation}
h_{\rm BG}^{2} = {1 \over \nu_{\rm obs}}\int h_{\rm BH}^{2} dR,
\end{equation}

\par\noindent (see de Araujo, Miranda \& Aguiar 2000 for
details).

The dimensionless amplitude produced by the collapse of a star, or
star cluster, to form a black hole is (Thorne 1987)

\begin{eqnarray}
\lefteqn{h_{\rm BH}=\bigg({15\over 2{\rm \pi}}\varepsilon_{\rm GW
}\bigg)^{1/2}{G\over c^{2}} {M_{\rm r}\over r_{\rm z}} {} }
\nonumber\\ & & {} \simeq 7.4\times 10^{-20}\varepsilon_{\rm GW
}^{1/2}\bigg({M_{\rm r}\over {\bf M}_{\odot}}\bigg) \bigg({r_{\rm
z}\over 1{\rm Mpc}}\bigg)^{-1}, \nonumber \\ \label{hBH}
\end{eqnarray}

\noindent where $\varepsilon_{\rm GW}$ is the efficiency of
generation of GWs and $r_{\rm z}$ is the distance to the source.

The collapse of a star to a black hole produces a signal with
frequency (Thorne 1987)

\begin{eqnarray}
\lefteqn{\nu_{\rm{obs}} = {1\over 5{\rm \pi} M_{\rm r}}{c^{3}\over
G}(1+z)^{-1} {} } \nonumber\\ & & {} \simeq 1.3\times 10^{4}{\rm
Hz}\bigg({{\rm M}_{\odot}\over M_{\rm r}}\bigg)(1+z)^{-1},
\end{eqnarray}

\noindent where the factor $(1+z)^{-1}$ takes into account the
redshift effect on the emission frequency, that is, a signal
emitted at frequency $\nu_{\rm e}$ at redshift $z$ is observed at
frequency $\nu_{\rm{obs}}=\nu_{\rm e}(1+z)^{-1}$. The observed
signal is in the range

\begin{equation}
{1.04\times 10^{3}\over (1+z_{\rm ci})} \;{\rm Hz} \leq
{\nu_{\rm{obs}}} \leq {5.2\times 10^{3}\over (1+z_{\rm cf})}\;{\rm
Hz}, \label{freq}
\end{equation}

\noindent obtained using the mass upper limit $m_{\rm u} = 125\,
{\rm M}_\odot$, the mass lower limit $m_{\rm min}=25\, {\rm
M}_\odot$, and $\alpha=0.1$.

For the differential rate of black hole formation we have

\begin{equation}
dR_{\rm BH} = \dot\rho_{\star}(z) {dV\over dz} \phi(m)dmdz,
\end{equation}

\noindent where $\dot\rho_{\star}(z)$ is the star formation rate
(SFR) density (in ${\rm M}_{\odot}\,{\rm yr}^{-1}\,{\rm
Mpc}^{-3}$) and $dV$ is the comoving volume element.

From the above equations we obtain for the dimensionless amplitude

\begin{eqnarray}
h_{\rm BG}^{2} &=& {(7.4\times 10^{-20}\alpha)^{2}\varepsilon_{\rm
GW} \over \nu_{\rm{obs}}} \nonumber\\ && \times \bigg[\int_{z_{\rm
cf}}^{z_{\rm ci}} \int_{m_{\rm min}}^{m_{\rm u}}\bigg({m\over {\rm
M}_{\odot}}\bigg)^{2}\bigg({d_{\rm L}\over 1{\rm Mpc}}\bigg)^{-2}
   \dot\rho_{\star}(z) \nonumber\\ && \times{dV\over
dz} \phi(m)dmdz\bigg].\label{hBG}
\end{eqnarray}

In equation (\ref{hBG}) $d_{\rm L}$ is the luminosity distance to
the source. The comoving volume element is given by

\begin{equation}
dV = 4{\rm \pi}\bigg({c\over H_{0}}\bigg) r_{\rm z}^{2}
{\mathcal{F}}(\Omega_{\rm M},\Omega_{\Lambda},z) dz,
\end{equation}

\par\noindent with

\begin{equation}
{\mathcal{F}}(\Omega_{\rm M},\Omega_{\Lambda},z) \equiv {1\over
\sqrt{(1+z)^2(1+\Omega_{\rm M}z)-z(2+z)\Omega_{\Lambda}}},
\end{equation}

\noindent and the comoving distance, $r_{\rm z}$, is

\begin{equation}
r_{\rm z}={c\over H_{0}\sqrt{|\Omega_{\rm k}|}}S\bigg(
\sqrt{|\Omega_{\rm k}|}\int_{0}^{z}{dz'\over {\mathcal{F}}(\Omega_{\rm
M},\Omega_{\Lambda},z')}\bigg),
\end{equation}

\par\noindent where

\begin{equation}
\Omega_{\rm M}=\Omega_{\rm DM}+\Omega_{\rm B} \qquad {\rm and}
\qquad 1 = \Omega_{\rm k}+\Omega_{\rm M}+\Omega_{\Lambda}
\end{equation}

\par\noindent are the usual density parameters for the
matter (M), i.e., dark matter (DM) plus baryonic matter (B),
curvature (k) and cosmological constant ($\Lambda$). The function
S is given by

\begin{eqnarray}
  S(x) = \cases { \sin x  &if  closed, \cr x &if  flat, \cr \sinh x &if  open.}
\end{eqnarray}

\noindent The comoving distance is related to the luminosity
distance by

\begin{equation}
d_{\rm L} = r_{\rm z} (1+z).
\end{equation}

The set of equations presented above can be used to find the
dimensionless amplitude of the GW background generated by black
hole formation as a function of the SFR density, and related to
the `first light' epoch.

It is worth mentioning  that the formulation used here is similar
to that used by Ferrari et al. (1999a), but instead of using an
average energy flux taken from Stark \& Piran (1986), who
simulated the axisymmetric collapse of a rotating polytropic star
to a black hole, we use equation (\ref{hBH}) to obtain the energy
flux, which takes into account the most relevant quasi-normal
modes of a rotating black hole and represents a kind of average
over the rotational parameter (see de Araujo et al. 2000). Both
formulations present similar results, since in the end the most
important contributions to the energy flux come from the
quasi-normal modes of the black holes formed, which account for
most of the gravitational radiation produced during the collapse
process.

The SFR density, however, for the formation of the first stars is
unknown. Star formation in other media could in principle give us
some information on how things occurred in the `first light'
epoch, but unfortunately as one can see in what follows there are
no compelling arguments in this direction.

The formation of a bound cluster of stars requires a star
formation efficiency of $\sim 50$ per cent when the cloud
disruption is sudden and $\sim 20$ per cent when cloud disruption
takes place on a longer time-scale (Margulis \& Lada 1983; Mathieu
1983; Ciardi et al. 2000). We define `star formation efficiency'
as the fraction of gas of a cloud that is converted into stars
[this definition is similar to that used by Ciardi et al. (2000),
among others].

On the other hand, Pandey, Paliwal \& Mahra (1990) have
investigated the influence of the IMF on the star formation
efficiency, for clouds of different masses, and have concluded
that the efficiency decreases if massive stars (the most
destructive ones) are formed earlier. Some studies have also
analysed the molecular gas properties and star formation in nearby
nuclear starburst galaxies (see, e.g., Planesas, Colina \&
Perez-Olea 1997) indicating the existence of giant molecular
clouds with masses $\sim 10^8-10^9 {\rm M}_\odot$, in which star
formation process occurs in a short time ($< 3\times 10^7\,{\rm
yr}$) with efficiency of conversion of gas into stars ${^<_\sim}\,
10$ per cent. All these studies show the large uncertainties in
the star formation efficiency.

Probably, the best way to infer the SFR density is to relate it to
studies concerning the reionization of the Universe. In the
introduction of the present paper we argued that there are
compelling arguments in favor of a reionization phase of the
Universe, which probably occurred at redshifts in the range $6 < z
< 30$ (Tegmark, Silk \& Blanchard 1994; Venkatesan 2000). There
are at least two compelling reasons for reionization through
Population III stars to be considered. First, the Population III
stars are expected to form at $z\, ^>_\sim 10$, being capable of
ionizing hydrogen. Secondly, the first stars  create heavy
elements, and can account for the metalicity of $\sim
10^{-2}Z_{\odot}$ found in ${\rm Ly}-\alpha$ forest clouds
(Venkatesan 2000). It is found that the amount of baryons
necessary to participate in early star formation, to account for
the reionization, would amount to a small fraction, $f_{\star}$,
of all baryons of the Universe (see, e.g., Venkatesan 2000; Loeb
\& Barkana 2001).

The above discussion suggests that we can write the SFR density as
follows:

\begin{equation}
\dot\rho_\star \equiv {d\rho_{\star}\over dt}= {d\over dt}
[\Omega_{\star}\;\rho_{\rm c}\;(1+z)^{3}] ,
\end{equation}

\par\noindent where the term in brackets
represents the stellar mass density at redshift $z$, with
$\rho_{\rm c}$ the present critical density and $\Omega_{\star}$
the stellar density parameter. The latter can be written  as a
fraction of the baryonic density parameter, namely,
$\Omega_{\star}=f_{\star}\Omega_{\rm B}$.

Another relevant physical quantity associated with the GW
background, produced by the first stars, is the closure energy
density per logarithmic frequency span, which is given by

\begin{equation}
\Omega_{\rm GW} = {1\over \rho_{\rm c}} {d\rho_{\rm GW}\over d\log
\nu_{\rm{obs}}}.
\end{equation}

\noindent The above equation can be rewritten as

\begin{equation}
\Omega_{\rm GW} = {\nu_{\rm{obs}}\over c^{3}\rho_{\rm c}}F_{\nu} =
{4{\rm \pi}^{2}\over 3H^{2}_{0}}\nu_{\rm{obs}}^{2} h_{\rm BG}^{2}.
\end{equation}

In the next section we present the numerical results and
discussions, which come mainly from equation (\ref{hBG}). Looking
at this equation one notes that, to integrate it, one needs to
choose the IMF, and to set values for the following parameters:
$z_{\rm ci}$, $z_{\rm cf}$, $\alpha$, $\varepsilon_{\rm GW}$,
$f_{\star}$, $H_{0}$, $\Omega_{\rm B}$, $\Omega_{\rm M}$ and
$\Omega_{\Lambda}$.

\section{NUMERICAL RESULTS AND DISCUSSIONS}

Based on models of structure formation, the first objects that
collapsed should have had masses around $10^{6}{\rm M}_\odot$
(see, e.g., Tegmark et al. 1997, among others), and objects with
higher masses should have collapsed subsequently. Evidently, as
the density fluctuations could have had peaks higher than
$1\sigma$ values, clouds could have collapsed earlier. As a
result, there were clouds with different masses collapsing around
the redshift of collapse of $10^{6}{\rm M}_\odot$.

We have considered that the first stellar formation is related to
the collapse of Population III objects. To evaluate the GW
background produced by the formation of the Population III black
holes, it is necessary to know the redshifts at which they began
and finished being formed. This is a very hard question to answer,
since it involves knowledge of the role of the negative and
positive feedbacks of star formation, which are regulated by
cooling and injection of energy processes.

Should the stochastic GW background studied here be significantly
produced and detected at a reasonable confidence level, the
present study can be used to obtain the redshift range where the
Population III black holes were formed, independently of any CDM
modelling. In Fig. 1 an example is given of how one could get
$z_{\rm ci}$ and $z_{\rm cf}$ from the curve $h_{\rm BG}$ {\rm
versus} $\nu_{\rm obs}$. Knowing the frequency band $\nu_{\rm
min}-\nu_{\rm max}$ and using equation (\ref{freq}), one obtains
$z_{\rm ci}$ and $z_{\rm cf}$ (see Fig. 1), which are, therefore
observable. Note that we have assumed as did Ferrari et al.
(1999a,b) that $\alpha$ is a constant ($\alpha =0.1$).

\begin{figure}
\begin{center}
\leavevmode
\centerline{\epsfig{figure=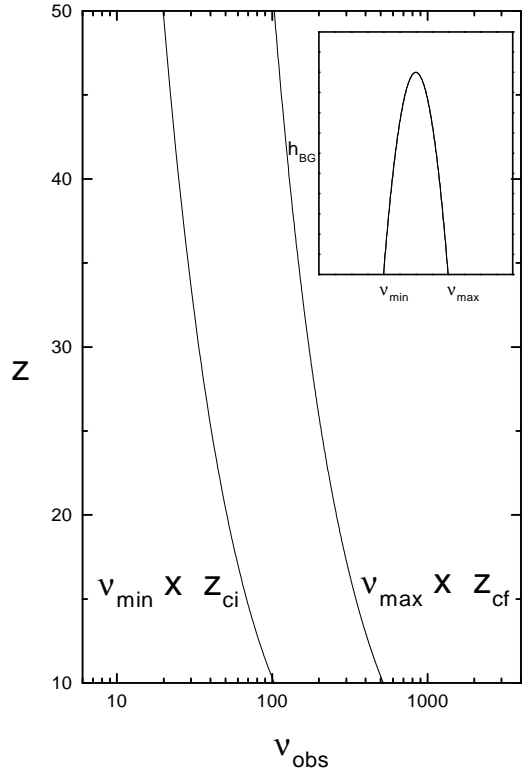,angle=360,height=12cm,width=8.5cm}}
\caption{Example of how one could obtain from $h_{\rm BG}$ {\rm
versus} $\nu_{\rm obs}$ the initial (final) redshift  $z_{\rm ci}$
($z_{\rm cf}$) of the GW emission period. We have adopted
$\alpha=0.1$.}
\end{center}
\end{figure}

Is is worth noting that $\alpha$ may depend sensitively on the
metallicity: the lower the value of $Z$, the higher are the
remnant masses and the less ejected material there is relative to
$Z_{\odot}$ stars. More realistically there would be a dependence
of $\alpha$ on the progenitor mass. On the other hand, the value
$\alpha=0.1$ adopted can be considered as a mean value for the
progenitor masses studied here. If $\alpha$ is not very well
determined, this would mean that the observed frequency band does
not uniquely fix the redshift band where the black holes are
formed.

In the introduction of the present paper we mention that different
studies related to the reionization of the Universe set this epoch
as being somewhere in the range $6 < z < 30$. It is also clear
that the reionization process is not instantaneous. Stars start
forming at different redshifts, creating ionized bubbles
(Str\"{o}mgren spheres) around themselves, which expand into the
intergalactic medium (IGM), at a rate dictated by the source
luminosity and the background IGM density (Loeb \& Barkana 2001).
The reionization is complete when the bubbles overlap to fill the
entire Universe. Thus the epoch of reionization is not the epoch
of star formation. There is a non-negligible time-span between
them. Here, we have chosen different formation epochs to see their
influence on the putative GW background and also to see if it
could be detected by the forthcoming GW antennas.

The first relevant quantity appearing in the equation for the GW
background is the IMF. As discussed in the previous section, we
have adopted the standard IMF ($x=1.35$), as our fiducial case,
and have also studied two other cases to assess the role of the
IMF variations on $h_{\rm BG}$.

The second relevant parameter is $\varepsilon_{\rm GW}$, the
efficiency of production of GWs, whose distribution function is
unknown. Thus, we have parameterized our results in terms of its
maximum value, namely, $\varepsilon_{\rm GW_{\rm max}}=7 \times
10^{-4}$, which is obtained from studies by Stark \& Piran (1986)
who simulated the axisymmetric collapse of a rotating star to a
black hole. We will see below that, if $\varepsilon_{\rm GW}$ is a
very tiny fraction of the maximum value, the detection of the GW
background whose existence we propose, is very improbable, even
for advanced antennas.

To calculate $h_{\rm BG}$ we still need to know $\Omega_{\star}$,
which has a key role in the definition of the SFR density. As
discussed in the previous section studies related to the
reionization of the Universe can shed some light on
$\Omega_{\star}$. From different studies one can conclude that a
few percent, maybe up to $\sim 10$ per cent, of the baryons must
be condensed into stars in order for the reionization of the
Universe to take place. Here we have set the value of
$\Omega_{\star}$ in such a way that it amounts to $1$ per cent of
all baryons (our fiducial value). In the next section we discuss
the detectability of the GW background, and we then parameterize
our results in terms of $f_{\star}=\Omega_{\star}/\Omega_{\rm B}$.

Looking at equation (\ref{hBG}) one could think it would depend
critically on the cosmological parameters: $H_{0}$, $\Omega_{\rm
B}$, $\Omega_{\rm DM}$, $\Omega_{\Lambda}$. However, our results
show that, given the redshifts involved in our calculations,
$h_{\rm BG}$ depends only on $H_{0}$ and $\Omega_{\rm B}$; the
latter dependence occurs because this parameter appears in the SFR
density. The quantity $h_{100}^{2}\Omega_{\rm B}=0.019\pm 0.0024$
(where $h_{100}$ is the Hubble parameter given in terms of 100
${\rm km\;s^{-1}\;Mpc^{-1}}$) is obtained from big bang
nucleosynthesis studies (see, e.g., Burles et al. 1999).

In Table 1 we present the redshift band, $z_{\rm ci}$ and $z_{\rm
cf}$ for the models studied and the corresponding GW frequency
bands. For the cosmological parameters we have adopted
$h_{100}=0.65$, $\Omega_{\rm M}=0.3$, $\Omega_{\rm B}=0.045$ and
$\Omega_{\Lambda}=0.7$. Keep in mind that our results are
sensitive to the combination $h_{100}^{2}\Omega_{\rm B}$. We have
also adopted $\alpha=0.1$, $f_{\star}=0.01$ and the standard IMF.

Note that no structure formation model has been used to find the
black hole formation epoch. Instead we have simply chosen the
values of $z$ to see whether it is possible to obtain detectable
GW signals. In the next section it will be seen that, unless
$\varepsilon_{\rm GW}$ is negligible, the GW background that we
propose here can be detected. Our choices, however, can be
understood as follows. The greater the redshift formation, the
more power the masses related to the Population III objects have.
Thus, of our models A to D, our model D (A) has more (less) power
when compared to the others. Models E, F and G would mean a more
extended star formation epoch, which means that the feedback
processes of star formation are such that they allow a more
extended star formation epoch when compared to models B, C and D,
respectively.

Concerning the reionization epoch, as already mentioned, it
occurred at lower redshifts as compared to the first star
formation redshifts. Loeb \& Barkana (2001) found, for example,
that if the stars were formed at $z \simeq 10-30$, with standard
IMF, they could have reionized the Universe at redshift $z \sim
6$. Our models A, B and E, for example, could account for such a
reionization redshift.

In addition, we also consider a model with $z_{\rm cf}=5$ (see
model H of Table 1) to verify if a final epoch of star formation
close to this redshift could produce a detectable signal for the
VIRGO and LIGO experiments.

If the process of structure formation of the Universe and the
consequent star formation were well known, one could obtain the
redshift formation epoch of the first stars. On the other hand, as
discussed below, if the GW background really exists and is
detected, one can obtain information about the formation epoch of
the first stars.

\begin{table}
\caption{The redshifts of collapse for our models and the
corresponding GW frequency bands. The cosmological parameter
$h_{100}^{2}\Omega_{\rm B}=0.019$ (see the text), $\alpha=0.1$,
$f_{\star}=0.01$ (our fiducial value) and the standard IMF are
adopted.}
\begin{flushleft}
\begin{tabular}{ccccccc}
\hline Model & $z_{\rm ci}$ & $z_{\rm cf}$ &  $ \Delta\nu $ \\
      &               &               &      (Hz)      \\
\hline
A  &  20  &  10 & 50-470 \\
B  &  30  &  20 & 34-250 \\
C  &  40  &  30 & 25-170 \\
D  &  50  &  40 & 20-130 \\
E  &  30  &  10 & 34-470 \\
F  &  40  &  10 & 25-470 \\
G  &  50  &  10 & 20-470 \\
H  &  15  &   5 & 65-870 \\
\hline
\end{tabular}
\end{flushleft}
\end{table}

A relevant question is whether the background we study here is
continuous or not. The duty cycle indicates if the collective
effect of the bursts of GWs generated during the collapse of a
progenitor star generates a continuous background. The duty cycle
is defined as follows:

\begin{equation}
DC = \int_{z_{\rm cf}}^{z_{\rm ci}} dR_{\rm
BH}\bar{\Delta\tau_{\rm GW}}(1+z),
\end{equation}

\noindent where $\bar{\Delta\tau_{\rm GW}}$ is the average time
duration of single bursts at the emission, which is inversely
proportional to the frequency of the lowest quasi-normal mode of
the rotating black holes (see, e.g., Ferrari et al. 1999a), which
amounts to $\sim 1\,{\rm ms}$ for the mass range of the black
holes considered here.

\begin{figure}
\begin{center}
\leavevmode
\centerline{\epsfig{figure=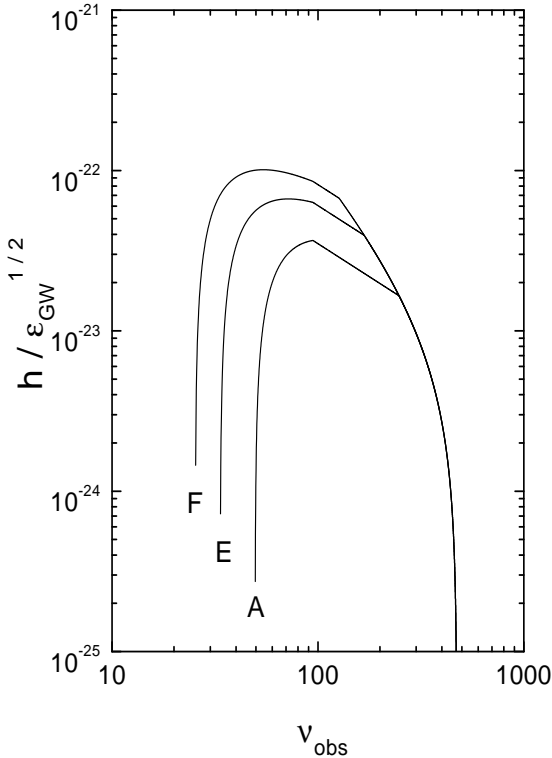,angle=360,height=12cm,width=8.5cm}}
\caption{The background amplitude of the GWs as a function of
$\nu_{\rm obs}$ and the efficiency of GW generation
$\varepsilon_{\rm GW}$ for models A, E and F of Table 1}
\end{center}
\end{figure}

Since the star formation rate could be high, a significant amount
of GWs could be produced. We also note that, independently of the
primordial cloud mass and of redshift of collapse, star formation
occurring at high redshift could produce high duty cycle values,
which lead us to conclude that the stochastic GW background could
be continuous. For all the models studied here the duty cycle is
$\gg 1$.

The amplitude $h_{\rm BG}$ of the GW background in terms of
$\varepsilon_{\rm GW}$ (the efficiency of generation of GWs), in
the frequency band $\nu_{\rm min}-\nu_{\rm max}$ is shown in Fig.
2 for the models A, E and F of Table 1. In the next section we
discuss the detectability of such a putative background.

Note that, the earlier that star formation occurs, the greater is
the GW amplitude $h_{\rm BG}$. This can be explained as follows.
For a given individual source, the higher the redshift, the lower
is the amplitude of the GWs generated. On the other hand, the
higher the redshift of star formation, the greater is the SFR
density. As a result there is a more significant overlapping of
bursts of GWs at higher redshifts.

We find, for example, that the formation of Population III of
black holes, in model D, could generate a stochastic GW background
with amplitude $h_{\rm BG} \simeq (0.8-2)\times 10^{-24}$ and a
corresponding closure density of $\Omega_{\rm{GW}}\simeq
(0.7-1.4)\times 10^{-8}$, in the frequency band $\nu_{\rm{obs}}
\simeq 20-130\, {\rm Hz}$ (assuming an efficiency of generation
$\varepsilon_{\rm GW} \simeq 7\times 10^{-4}$, the maximum one).

It could be argued that the formation of stars could not be
restricted to the epoch of the collapse of the first Population
III objects. A new surge of star formation associated with the
collapse of Population III objects of $M > 10^{6}{\rm M}_{\odot}$
at lower redshifts could occur. The existence of this new surge of
star formation would depend on the role of the negative and
positive feedbacks, which are regulated by cooling and injection
of energy processes of the previous star formation surge. If
another surge of star formation took place, then another GW
background could be generated, and a partial superposition with
the background previously generated could also occur. As a result
for some frequency bands the GW amplitude could be enhanced.

Another possibility would be a star formation process taking place
during the time of collapse of the first Population III objects
and continuing during the collapse of objects of higher masses. In
such a case the star formation would occur for a large redshift
span, and as a result the frequency band, the amplitude and the
closure energy density of GWs could be larger. As before, the role
of the negative and positive feedbacks of the star formation would
be the key point. In the models E, F and G we have considered such
a possibility (see also Fig. 2).

Certainly, the GW background produced depends on the star
formation history. A different star formation history would
produce different results for both the values of $h_{\rm BG}$ and
frequency bands. However, it is hard to avoid the conclusion that,
if the first stars had been formed at high redshift, a significant
amount of GWs would have been produced as well.

To assess the role of possible IMF variations we have considered,
besides the standard IMF, two others, namely, with $x=0.3$ and
$x=1.85$, which yield ten times and one-tenth of the mass fraction
of black holes of the standard IMF, respectively. For the model A,
our calculations show that for $x=0.3$ ($x=1.85$) the maximum
$h_{\rm BG}$ is a factor $\simeq 3$ ($\simeq 4$) greater (smaller)
than case with standard IMF ($x=1.35$). In the next section we
consider the role of the IMF variations on detectability of the
background of the GW background that we propose exists.

\section{DETECTABILITY OF THE BACKGROUND OF GRAVITATIONAL WAVES}

The background predicted in the present study cannot be detected
by single forthcoming interferometric detectors, such as VIRGO and
LIGO (even by advanced ones). However, it is possible to correlate
the signal of two or more detectors to detect the background that
we propose exists. Michelson (1987) was the first to show that
this kind of signal can, in principle, be detected by correlating
the outputs of two different detectors. However, the main
requirement that must be fulfilled is that they must have
independent noise. This study was improved by Christensen (1992)
and by Flanagan (1993). The reader should also refer to the papers
by Allen (1997) and Allen \& Romano (1999) who also deal in detail
with such an issue.

To assess the detectability of a GW signal, one must evaluate the
signal-to-noise ratio (S/N), which for a pair of interferometers
is given by (see, e.g., Flanagan 1993; Allen 1997)

\begin{equation}
{\rm (S/N)}^2=\left[\left(\frac{9 H_0^4}{50\pi^4} \right) T
\int_0^\infty d\nu \frac{\gamma^2(\nu)\Omega^2_{GW}(\nu) } {\nu^6
S_h^{(1)}(\nu) S_h^{(2)}(\nu)} \right]
\end{equation}

\noindent where $ S_h^{(i)}$ is the spectral noise density, $T$ is
the integration time and $\gamma(\nu)$ is the overlap reduction
function, which depends on the relative positions and orientations
of the two interferometers. For the $\gamma(\nu)$ function we
refer the reader to Flanagan (1993), who was the first to
calculate a closed form for the LIGO observatories. Flanagan
(1993; see also Allen 1997) showed that the best window for
detecting a signal is $0< \nu < 64$ Hz, where the overlap
reduction function has the greatest magnitude.

Here we consider, in particular, the LIGO interferometers. Their
spectral noise densities have been taken from a paper by Owen et
al. (1998)- who in turn obtained them from Thorne, by means of
private communication.

In Table 2 we present the S/N for one year of observation with
$\alpha=0.1$, $\Omega_{\rm B}h^{2}_{100}=0.019$, $f_{\star}=0.01$
and $\varepsilon_{\rm GW_{\rm max}}=7\times 10^{-4}$ for the
models of Table 1, for the three LIGO interferometer
configurations.

\begin{table}
\caption {For the models of Table 1 we present the S/N for pairs
of LIGO I, II and III (`first', `enhanced' and `advanced',
respectively) observatories for one year of observation. Note that
an efficiency of generation $\varepsilon_{\rm GW_{\rm max}} =
7\times 10^{-4}$ is assumed.}
\begin{tabular}{cccc}
\hline
Model &        &  S/N    &          \\
      & LIGO I  & LIGO II  & LIGO III \\
\hline
A  & $8.3\times 10^{-3}$  & $1.6$               & $6.6$  \\
B  & $8.5\times 10^{-3}$  & $2.3$               & $26 $  \\
C  & $8.7\times 10^{-3}$  & $2.7$               & $47 $  \\
D  & $8.1\times 10^{-3}$  & $2.5$               & $51 $  \\
E  & $2.7\times 10^{-3}$  & $5.7$               & $37 $  \\
F  & $5.0\times 10^{-3}$  & $12 $               & $120$  \\
G  & $7.7\times 10^{-2}$  & $21 $               & $260$  \\
H  & $4.6\times 10^{-3}$  & $0.5$               & $1.7$  \\
\hline
\end{tabular}
\end{table}

Note that for the `initial' LIGO (LIGO I) there is no hope of
detecting the GW background we propose here, even for ideal
orientation and locations of the interferometers, i.e.,
$|\gamma(\nu)|=1$. For the `enhanced' LIGO (LIGO II) there is some
possibility of detecting the background, since ${\rm S/N}
> 1$, if $\varepsilon_{\rm GW}$ is around the maximum value. Even
if the LIGO II interferometers cannot detect such a background, it
will be possible to constrain the efficiency of GW production.

The prospect for the detection with the `advanced' LIGO (LIGO III)
interferometers is much more optimistic, since the S/N for almost
all models is significantly greater than unity. Only if the value
of $\varepsilon_{\rm GW}$ were significantly lower than the
maximum value would the detection not be possible. In fact, the
S/N is critically dependent on this parameter, whose distribution
function is unknown.

Note, for example, that it is possible to detect a GW background
with the `advanced' LIGO, even for star formation for which
$z_{\rm cf}\sim 5$ (model H), if $\varepsilon_{\rm GW}$ is around
the maximum value.

Let us now look at how the variations of the parameters modify our
results. First of all, note that the larger the star formation
redshift band, the greater is the S/N. Secondly, the earlier the
star formation, the greater is the S/N. It is worth recalling
that, if one can obtain the curve of $h_{\rm BG}$ versus $\nu_{\rm
obs}$ and if the value of $\alpha$ is known, one can find the
redshift of star formation.

The S/N is also sensitive to variations of $\alpha$. The larger
$\alpha$, the lower are the GW frequencies and the higher is
$h_{\rm BG}$, and since the best window for detection is around $0
< \nu < 64 {\rm Hz}$, the S/N is higher.

Even if $\alpha$ is not known beforehand, it is possible to impose
a constraint on its values, and also on the redshift star
formation epoch. For example, if one found from GW observations
that the GW frequency band were $40-200$ Hz, one would obtain
(using Equation 11) that $\alpha\simeq 0.1-0.4$ and $z_{\rm
f}\simeq 5-50$. On the other hand, if one knew that the star
formation redshift band were $z_{f}\simeq 10-30$, through some
model of structure formation or whatever observational data, and
the GW frequency band were known, say $40-200$ Hz, one would
obtain that $\alpha\simeq 0.1$.

It would be interesting to perform a study considering $\alpha$ as
a function of the progenitor mass, which would result in a more
realistic model. There are some studies in the literature
considering how the remnant mass depends on the progenitor (see,
e.g., Fryer \& Kalogera 2001), but we will not consider such an
issue here.

To assess the role of the variations in the IMF, we modify its
exponent $x$. For $x=0.3$ ($x=1.85$), the S/N is $\sim 10$ ($\sim
0.1$) times the S/N of the standard IMF. As expected for an IMF
biased toward high- (low-) mass stars, where one has a greater
(lower) amount of black holes, the S/N is greater (lower).

Note that the S/N for a given formation epoch, IMF and $\alpha$,
and for one year of observation, still presents a dependence on
$\Omega_{\rm B}h^{2}$, $f_{\star}$ and $\varepsilon_{\rm GW}$,
namely

\begin{equation}
{\rm S/N} \propto  \Omega_{\rm B}h^{2}f_{\star}\;\varepsilon_{\rm
GW}.
\end{equation}

\noindent The value of $\Omega_{\rm B}h^{2}$ is well constrained
by primordial nucleosynthesis studies. For $f_{\star}$ we have
adopted a value of 0.01, which is a very conservative choice. Note
that the value for this parameter can be obtained from studies
concerning the reionization of the Universe, which is very
difficult to model, but there is some agreement in the literature
(see, e.g, Gnedin 2000; Venkatesan 2000; Loeb \& Barkana 2001)
among the different models of the reionization of the Universe
which lead us to conclude that $f_{\star}$ could range from a few
up to 15 per cent.

For $\varepsilon_{\rm GW}$, the situation is more complicated
since its distribution function is unknown. We have adopted here
the maximum value as a reference, but if its actual value is much
less than this value the S/N could be lower than unity for all the
models studied here, even for a LIGO III pair. Let us think of
what occurs with other compact objects, namely, the neutron stars,
to see if we can learn something from them. Hot and rapidly
rotating neutron stars can lose angular momentum to gravitational
radiation via the so-called r-mode instability (Anderson 1998).
This could explain why all known young neutron stars are
relatively slow rotators. The black holes could have had a similar
history, i.e., they could have been formed rapidly rotating and
lost angular momentum to gravitation radiation via their
quasi-normal modes. If this was the case, the value of
$\varepsilon_{\rm GW}$ could be near the maximum one, or in the
worst case it could have a value to produce ${\rm S/N}
>1$ at least for a LIGO III pair.

In order to assess the values of $f_{\star}$ and $\varepsilon_{\rm
GW}$ that yield ${\rm S/N} > 1$, for a given formation epoch, IMF,
$\alpha$ and  $\Omega_{\rm B}h^{2}$, we present in Fig. 3 the
regions in the $(f_{\star},\varepsilon_{\rm GW})$ plane where S/N
could be greater than unity for a pair of LIGO III
interferometers. Note that unless $\varepsilon_{\rm GW}$ is very
small, S/N can be significantly greater than unity, indicating
that the background could in principle be detected in the near
future.

\begin{figure}
\begin{center}
\leavevmode
\centerline{\epsfig{figure=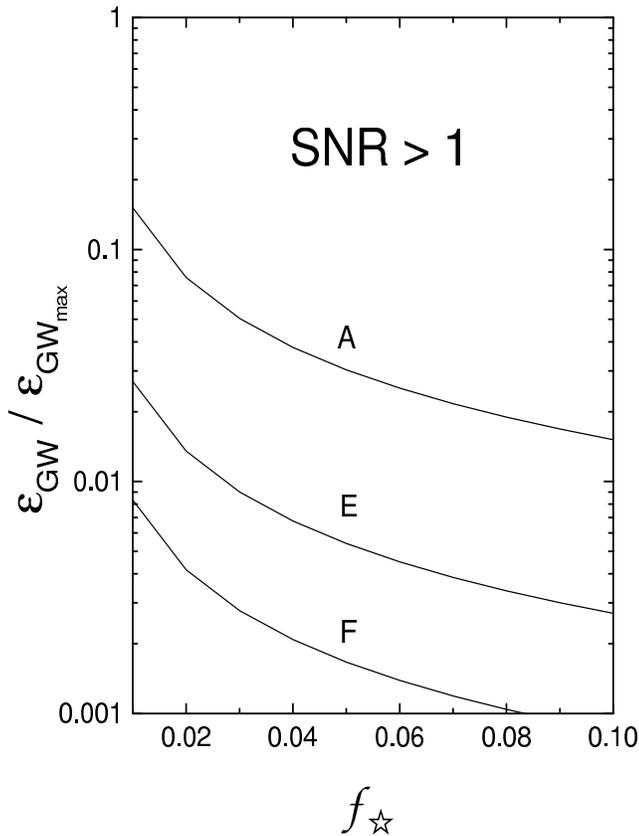,angle=360,height=12cm,width=8.5cm}}
\caption{Relative efficiency of GW generation, $\varepsilon_{\rm
GW }/\varepsilon_{\rm GW_{max}}$, as a function of the fraction of
baryons participating in the early star formation, $f_{\star}$,
for models A, E and F of Table 1 for a pair of LIGO III
interferometers. The curves represent where the ${\rm S/N}=1$;
above them ${\rm S/N} >1$. As in Table 2, $\varepsilon_{\rm
GW_{max}}~=~7\times~10^{-4}$ is assumed.}
\end{center}
\end{figure}

A relevant issue is whether there are other GW backgrounds that
could be confused with that of the present study. Relic GWs
generated in the very early Universe can in principle present a
signal in the LIGO bandwidth. The ordinary inflationary models,
however, predict $\Omega_{\rm GW} \sim 10^{-15}$ (see, e.g.,
Schultz 1999; Giovannini 2000; Maggiore 2000a,b). This is much
less than our models predict, and therefore undetectable even with
a pair of LIGO III interferometers. Other models, such as string
cosmologies, provide different predictions (see, e.g., Schutz
1999) with values of $\Omega_{\rm GW}$ that could be much greater
than our studies predict, which could render the background
studied here undetectable. Other GW backgrounds exist in the
bandwidth of LIGO, which could have been produced at $0< z < 5$,
namely: (a) a cosmological population of core-collapse supernovae
(Ferrari {et al. 1999a); (b) a population of young rapidly
rotating neutron stars (Owen et al. 1998, Ferrari et al. 1999b);
and (c) double neutron star binaries (Schneider et al. 2000). The
first two backgrounds, however, have energy density shifted for
higher GW frequencies when compared to our predictions. Moreover,
the S/Ns of these backgrounds are less than unity, even for a pair
of LIGO III. The last one has a frequency band ranging from $\sim
10^{-5}$ Hz up to $\sim 10^{2}$ Hz, and therefore there is a
partial overlap with the background of our study. The GW
amplitudes, however, would be comparable only if $\varepsilon_{\rm
GW}/\varepsilon_{\rm GW_{\rm max}} << 1$.

\section{CONCLUSIONS}

We present here a study concerning the generation of GWs produced
from a cosmological population of black holes. These objects are
formed as a consequence of the collapse of pre-galactic objects
(Population III objects) that form the first generation of stars
at high redshift. Our results show that different structure
formation models which predict the formation of the first objects
at $z > 10$ could, in principle, predict the formation of
pre-galactic black holes and a significant stochastic GW
background associated with them. Our results lead us to conclude
that star formation occurring at high redshifts  could have large
duty cycles and so the stochastic GW background generated is
continuous.

We consider that stars are formed following a Salpeter IMF and
having masses in the range $0.1-125 {\rm M}_\odot$. Certainly, the
results presented here are dependent on this particular choice. A
steeper IMF would modify the number of high-mass stars, modifying
the peak of $h_{\rm BG}$ and the frequency band of the GWs. For an
IMF with $x=0.30$ ($x=1.85$) the IMF is biased toward high- (low-)
mass stars, as a result the S/N is $\sim$10 ($\sim$0.1) times the
S/N predicted with the use of a standard IMF. It would be of
interest, however, to have a look in detail at studies of the
metallicity of high-z Ly $\alpha$ clouds to see if it is possible
to constrain the Population III IMF.

If we consider $\varepsilon_{\rm GW}\; \simeq\; 7.0\times 10^{-4}$
(see, e.g., Stark \& Piran 1986) then we obtain $h_{\rm BG} \simeq
(0.8-2)\times 10^{-24}$ and $\Omega_{\rm GW}\simeq (0.7-1.4)\times
10^{-8}$ at $\nu_{\rm{obs}}\simeq 20-130 \, {\rm Hz}$ for the
model D. Thus, this GW background produced as a consequence of the
formation of the first stars in the Universe is capable of being
detected by a pair of `advanced' LIGO interferometers.

As seen, with reasonable parameters, our results show that a
significant amount of GWs is produced related to the the
Population III black hole formation at high redshift, and can in
principle be detected by a pair of LIGO II (or most probably by a
pair of LIGO III) interferometers. However, a relevant question
should be considered: What astrophysical information can one
obtain from whether or not such a putative background is detected?

First, let us consider a non-detection of the GW background. The
critical parameter to be constrained here is $\varepsilon_{\rm
GW}$. A non-detection would mean that the efficiency of GWs during
the formation of black holes is not high enough. Another
possibility is that the first generation of stars is such that the
black holes formed had masses $> 100M_{\odot}$, and should they
form at $z > 10$ the GW frequency band would be out of the LIGO
frequency band.

Secondly, a detection of the background with a significant S/N
would permit us to obtain the curve $h_{\rm BG}$ versus $\nu_{\rm
obs}$. From it, as discussed above, one can constrain $\alpha$ and
the redshift formation epoch; and for a given IMF and $\Omega_{\rm
B}h^{2}$, one can also constrain the values of $f_{\star}$ and
$\varepsilon_{\rm GW}$. On the other hand, using the curve $h_{\rm
BG}$ versus $\nu_{\rm obs}$ and in addition other astrophysical
data, say CBR data, models of structure formation and reionization
of Universe, a constraint on $\varepsilon_{\rm GW}$ can also be
imposed.

\section*{Acknowledgments}
JCNA and ODM would like to thank the Brazilian agency FAPESP for
support (grants 97/06024-4, 97/13720-7, and 98/13735-7,
00/00116-9, 01/04086-0, respectively). ODA thanks CNPq (Brazil)
for partial financial support (grant 300619/92-8). We would also
like to thank the referee for helpful comments that we feel have
considerably improved the paper.

\label{lastpage}
\end{document}